\documentclass[preprint]{revtex4-2}
\usepackage[utf8]{inputenc}
\usepackage[english]{babel}
\usepackage{amsmath}
\usepackage{amsthm}
\usepackage{amsfonts}
\usepackage{amssymb}
\usepackage{graphicx}
\usepackage[hidelinks]{hyperref}
\usepackage{multirow}
\usepackage{tikz}
\usepackage{verbatim}
\usepackage[ruled,vlined]{algorithm2e}
\usepackage{enumitem}
\usepackage{float}

\usepackage{siunitx} 
\usepackage{isotope} 

\newcommand{\mylabel}[2]{#2\def\@currentlabel{#2}\label{#1}}

\newcommand{\bs}[1]{\boldsymbol{#1}}

\usepackage{color}

\newcommand{\zq}{\vartheta}

\newcommand{\zd}{\delta}

\newcommand{\zs}{\sigma}

\newcommand{\ze}{\varepsilon}

\newcommand{\zg}{\gamma}
\newcommand{\zl}{\lambda}

\newcommand{\zr}{\rho}
\newcommand{\za}{\alpha}

\newcommand{\zf}{\varphi}

\newcommand{\zh}{\eta}
\newcommand{\zn}{\nu}

\newcommand{\dif}{\; \textrm d }

\newcommand{\dst}[2]{\frac{  \dif^2 #1 }{  \dif #2^2   } }
\newcommand{\dpp}[2]{\frac{  \partial #1 }{  \partial #2   } }
\newcommand{\dpt}[2]{\frac{  \dif #1 }{  \dif #2   } }

\theoremstyle{definition}

\usepackage{csquotes}

\graphicspath{ {figure/} }

\begin{document}
\title{\bf Optimal control of geometric phase in pairs of interacting atoms traveling along two-dimensional closed paths.}
\author{Omar Morandi}
 \email{omar.morandi@unifi.it}
\affiliation{Department of Mathematics and Informatics 'Ulisse Dini', University of Florence, Viale Morgagni 67/A, 50134 Florence, Italy}
\begin{abstract}
    Universal quantum gates whose operation depends on the manipulation of the geometric phase of atomic systems are promising candidates for implementation of quantum computing. 
    We propose a scheme inducing a non-trivial Aharonov-Anandan geometric phase in pairs of atoms interacting via dipole-dipole potential. Our protocol relies on mobile optical trap technology and consists of steering a single atom along a closed loop. The trajectory of the atom is controlled by a mobile optical trap, and the shape of the path is designed by applying an optimal control procedure. The geometric phase is generated as a residual of the two-atom entanglement induced by the dipole-dipole interaction.  
    The stability of our scheme in the presence of noise or experimental imperfections is discussed.
	\end{abstract}
	    \maketitle
	\paragraph*{MSC:}{\small .....}

	\section{Introduction}\label{Sec_intro}
 	During the last decade, motivated by the great potentialities of quantum computing technology, geometrical and topological properties of quantum systems have gained growing interest. Geometrical descriptions of quantum systems find applications, for example, in modeling topological insulators, explaining topological protection, and designing synthetic gauge fields. 
	The geometrical representation of a quantum process emerges in a natural way when the Hilbert space of the system is expressed as a fiber bundle with non-trivial holonomy. 
    The geometrical and topological description of quantum systems with parametric control is generally associated with the concept of the geometric phase. By total phase one designs the global phase acquired by the system when the control parameters undergo to closed paths. Whenever such a phase is influenced only by the shape of these paths, the global phase is denoted by geometrical phase. 
	Pioneering work in this field was done by Pancharatnam in the context of the polarization of light \cite{Pancharatnam_56}. Successively, the concept of the adiabatic geometric phase was formalized by Berry. He observed that adiabatic cyclic parallel transport of quantum states causes the appearance of a well-defined quantum phase \cite{Berry_84,Barry_83}. The Berry phase is one of the fundamental breakthrough discoveries concerning non-trivial geometrical properties of quantum systems. 	
	The seminal result of Berry was rapidly extended to fiber bundles in which the geometric phase generalizes to the holonomy of non-adiabatic parallel transport and to non-Abelian gauge fields \cite{Wilczek_84,Aharonov_87,Anandan_88}.   
	Nowadays, Holonomic Quantum Control (HQC) techniques emerge as a powerful tool to design fault tolerant protocols applied to quantum information processing. The holonomic phase depends on the global properties of the controlled system and is less sensitive to the evolution details \cite{Zanardi_99}. A recent review of geometrical methods and applications to holonomic control is given by \cite{Zhang_23}. HQC protocols have been implemented in many systems, such as nitrogen-vacancy centers in diamonds \cite{Liu_23,Nagata_18,Sekiguchi_17,Zhou_17,Camejo_14}, ion traps \cite{Ai_20}, nuclear magnetic resonance systems \cite{Feng_13}, superconducting qubit \cite{Abdumalikov_13,Zhu_19}, quantum dots \cite{Greilich_09} and ultracold gases \cite{Ruseckas_05,Lin_11,Aidelsburger_15,Nguyen_15,Wu_16}. 
	
	The modern technology of optical traps for ultracold gases, opens the possibility to control the quantum dynamics of single or interacting atoms excited to Rydberg states.  
	Two-dimensional unstructured arrays of optically controlled neutral atoms have emerged as promising platforms for implementing quantum computing or studying fundamental interactions between atom pairs as, for example, van der Walls and dipolar potentials. Atoms at a temperature below to micro Kelvin can be trapped in optical lattices or in arrays of microscopic dipole traps. The trap configuration is almost fully controllable by optical addressing techniques \cite{Couvert_08,Chen_11,Rosi_13,Zhang_15,Gieseler_21,Hwang_23,Cicali_24,Morandi_25_arxive}. A review of recent advances in the manipulation of neutral atoms can be found in \cite{Browaeys_20}.
	Atomic states with large principal quantum numbers denoted as Rydberg states are particularly attractive for the studies of few- and many-body physics and for quantum information applications. Rydberg states are characterized by a long lifetime (which scales as the third power of the principal quantum number) and exhibit large tunable dipole moments. This leads to large van der Waals or dipole-dipole interaction strengths, corresponding to MHz frequencies \cite{Jaksch_00,Beguin_13,Li_14,Thaicharoen_15,Browaeys_16,Weber_17}. Rydberg state interactions hold unique potential for the implementation of quantum gates or encoding quantum information in spatially separated neutral atoms \cite{Jaksch_00,Saffman_10,Weimer_10,Petrosyan_17,Saffman_20,Liang_22,Evered_23} .
	
	Dipolar interactions between Rydberg states including anisotropy effects have been directly measured \cite{Leseleuc_17}. Symmetry protected topological phases in one-dimensional chains of Rydberg atoms under dipolar interaction have been observed in \isotope[87]{Rb} gases \cite{Leseleuc_19}. Rydberg states have also been deeply investigated from a theoretical point of view. Open source libraries are now available to compute the dipole-dipole interaction strengths for many atoms \cite{Sibalic_17}. 
    	
	Reminiscent of the Berry phase approach, the majority of the early schemes of HQC were based on adiabatic hypothesis. In practical terms, the adiabatic hypothesis requires that the variations of parameters controlling the system should be slow enough to ensure that the system is maintained in the desired Hilbert subspace. For this reason, adiabatic evolution limits the speed of the gate implementations and, in general, it requires long time to achieve goals. 
	To improve the performances of geometric quantum computation, Nonadiabatic Holonomic Quantum Computation (NHQC) schemes have been investigated, and nowadays are promising candidates for implementing robust and high-fidelity phase control of quantum gates.	
	According to the Anandan definition of the Abelian or non-Abelian geometric phase, every subset of the eigenstates of the quantum evolution operator may be used as a set of cyclic states on which nonadiabatic geometric quantum gates can be implemented. Inside such subspaces, the holonomy and parallel transport are defined in a natural way. 
	Various schemes to implement NHQC have been proposed \cite{Sjoqvist_12,Liu_19,Kang_20,Yao_24}, and have been verified experimentally \cite{Xu_12,Feng_13,Kang_18,Xu_18,Zhu_19,Zhao_19,Ai_20}.
	%

Ideally, holonomic control should avoid the presence of a residual dynamical phase at the end of the protocols. In many cases, this operation can be challenging and specific correctors are designed to achieve dynamic phase rejection.  
    In few cases it is possible to reject the dynamical contribution to the phase by exploiting global symmetries of the systems \cite{Yao_24}, or tailoring loops associated to zero or multiple value of $2 \pi$ of dynamic phase \cite{Zhang_23}.
    
	Recently, in \cite{Song_24} the possibility of inducing holonomy in Cesium Rydberg atoms using the dipole interaction has been proposed. The two-particle dipole-dipole interaction is controlled by trapping two single atoms via optical tweezers. The atom pair is maintained at a fixed interatomic distance. The internal state of the atoms is controlled by optical fields whose phase is engineered to achieve robust logical gates. 
	
	In this contribution, we propose a scheme for implementing a geometric phase in ultracold atom systems based on moving optical trap technology. The geometric phase and the two-particle entanglement originate from the dipolar interaction of the atom pair. Unlike previous schemes, in which the atoms lie in fixed positions during the interaction \cite{Jaksch_00,Wilk_10,Feng_13,Leseleuc_17,Petrosyan_17,Levine_18,Kang_20,Ai_20,Song_24}  in our protocol, the geometric phase is controlled by moving the atoms along a two-dimensional closed-loop trajectory.

	

\section{Controlling Aharonov Anandan phase in pairs of interacting Rydberg atoms}\label{Sec_mod_intro}
	
We describe a protocol whose goal is to induce a geometric phase in Rydberg atom pairs by controlling the dipole-dipole interaction. The geometric phase is a manifestation of the two-particle correlation induced by the dipole potential. We focus on a single pair of atoms and we apply the Aharonov-Anandan (AA) definition of Abelian geometric phase \cite{Aharonov_87}.  In our scheme, the AA phase is generated by steering one atom along a designed closed trajectory. The position of the atom is controlled by a mobile optical tweezer. The second atom is held in a fixed position. During the evolution, the two particles become entangled and a non-trivial AA phase results. 

In order to set the notation, we introduce the definition of the Abelian geometric phase proposed by Aharonov and Anandan. 
We consider two atoms described by the two-atom wave function $\Psi\in \mathbb{H}_{2p}\doteq \mathbb{C}^d\otimes \mathbb{C}^d$, where $\mathbb{H}_{2p}$ denotes the two-particle Hilbert space of dimension $d^2$.  We assume that the atoms are well localized so that their position can be safely described by the classical position-momentum coordinates. This assumption applies to typical experimental setups of ultracold atoms in optical traps \cite{Gieseler_21,Hwang_23,Morandi_25_arxive}. We denote by $(\mathbf{r}^1(t),\mathbf{r}^2(t))$ the time-dependent coordinates of the atoms. The quantum dynamics of the two atoms is governed by Schr\"odinger equation $i\hbar\dpp{\Psi}{t}= \mathcal{H}_{dd} (\mathbf{r}^1(t),\mathbf{r}^2(t))\Psi$, where $\mathcal{H}_{dd}(\mathbf{r}^1,\mathbf{r}^2)$ denotes the dipole-dipole Hamiltonian parametrized by the atom positions. The detailed expression of the Hamiltonian is described in Sec. \ref{sec_model}. 
According to the AA approach, we are interested in the cases in which the two-particle wave function undergoes a periodic evolution after the first atom has completed a closed loop. We denote by $T$ the corresponding time interval. As an alternative to the solution of the Schr\"odinger equation, the wave function $\Psi(T)$ can be obtained by evaluating the quantum propagator $U\in C^1([0,T], \mathbb{U}_t(\mathbb{H}_{2p}) )$,  where $\mathbb{U}_t$ denotes the family of strongly continuous unitary semigroups parametrized by the time. Similarly to the wave function, the evolution of the propagator is obtained from the Schr\"odinger equation $i\hbar \dpp{U}{t}=\mathcal{H}_{dd}(\mathbf{r}^1(t),\mathbf{r}^2(t))\;  U$, with the initial condition $U(0)=\zd$, where $\zd$ denotes the Kronecker delta. The trajectories of the atoms $(\mathbf{r}^1(t),\mathbf{r}^2(t))$ are influenced by the total force fields experienced by the atom pair, and in general cannot be assigned at will. Concerning ultracold atoms trapped by optical tweezers, up to a good degree of approximation, we can limit ourselves to consider two relevant force fields: the net force generated by the optical tweezers which can be controlled at some degree, and the dipole-dipole force. The dipole-dipole force depends on the quantum wave function itself. In general, its contribution cannot be neglected and may sensibly modify the particle trajectories. As described in Sec. \ref{sec_model}, in our model the dipolar force is evaluated selfconsistently, and the particle trajectories are determined by a nonlinear optimal control procedure. In order to illustrate our method, at the present stage we discard the details on how to fix the external field to steer the particles along the desired trajectories and assume that the particle motion is known. With the use of the propagator, the two-atom wave function is given by $\Psi(t)=U(t)\Psi(0)$, where $\Psi(0)$ denotes the initial condition. According to Aharonov and Anandan, the difference between each eigenvalue of the propagator $U(T)$ and the associated integral of the energy in the time interval $T$, can be interpreted as a well-defined geometric phase related to the periodic motion of the atoms. The eigenvalue equation for the propagator writes $U(T)\Phi_n=e^{i\zg_n }\Phi_n$ with $n=1,\ldots,d$. The phase $\zg_n$ associated with the $n$-th eigenvalue is denoted by the total phase. We assume that the eigenvectors $\Phi_n$ are non degenerate, which is the relevant case for the dipole-dipole Hamiltonian. By construction, the solutions of the Schr\"odinger equation whose initial condition coincides with one eigenvector $\Psi_n(0)=\Phi_n$, form a set of periodic functions as a ray in the Hilbert space, i. e. $\Psi_n(T)=e^{i\zg }\Psi_n (0)$. The AA geometric phase $\zg_n^g\doteq \zg_n - \zg_n^d$ is obtained as the difference of the total phase with the dynamical phase $\zg^g_n$, where $\zg^d_n=-  \int_0^T\langle \Psi_n(t) |\mathcal{H}_{dd}| \Psi_n (t)\rangle \dif t$. 

The application of the AA definition of geometric phase to systems of atom pairs interacting via dipole-dipole Hamiltonian encounters few difficulties. 
The eigenvectors $\Phi_n$ obtained by diagonalizing the propagation operator, are truly two-particle states i. e. cannot be expressed as tensor product of two single-particle wave functions, $\Phi_n\neq \zh^1 \otimes \zh^2$  for any $\zh^1,\zh^2\in \mathbb{H}_{sp}\doteq\mathbb{C}^d$. The family of eigenvectors $\{\Phi_i\}$ contains all possible initial states of the two-atom system. Preparing highly correlated states described by non-separable two-particle wave functions is extremely challenging and may constitute a serious limitation to the experimental implementation of our procedure. 
For this reason, one of the objectives of our work is to derive a protocol implementing AA geometric phase which ensures that the initial and final two-particle states can be expressed as the tensor product of two single-particle states. This is achieved by designing an optimization procedure that can select the trajectories associated with factorized states. We define a separability function $F(\Psi)$, quantifying the degree of separability of the solution as follows. We denote by $\zr_{2p} \doteq |\Psi \rangle \langle \Psi|$ the two-particle density matrix associated with the pure state $\Psi$. The single particle density matrices $\zr_1\doteq \mathrm{tr}_2(\zr_{2p}) $ and $\zr_2\doteq \mathrm{tr}_1(\zr_{2p}) $ associated with the first and second particle, respectively, are obtained by projecting the two-particle density matrix onto the single-particle Hilbert spaces of each particle. Here, $\mathrm{tr}_a$ denotes the partial trace over the degrees of freedom of the $a-$th particle. In general, as a consequence of the projection procedure, the single-particle density matrices are associated to mixed states. The two-particle wave function is factorizable if and only if both $\zr_a$ are single-particle density matrices associated with pure states. The maximum value of the eigenvalues provides a simple indication of the ``purity'' of a density matrix. We denote by $\zh_1$ and $\zh_2$ the single particle eigenvectors associated with the highest eigenvalue of the matrices $\zr_1$ and $\zr_2$, respectively. In our algorithm, a convenient measure of the single-particle purity associated with the decomposition of $\Psi$ is expressed by the quantity $F(\Psi)\doteq  |\langle\Psi|\zh_1 \otimes \zh_2 \rangle | $. Clearly $0\leq F\leq 1$, and $F=1$ if and only if $\Psi =e^{i\zf} \left(\zh_1 \otimes \zh_2\right)$, where $\zf $ is an irrelevant phase factor. 

\subsection{Optimal control of the atom trajectories}\label{sec_model}
In this section, we describe the mathematical formulation of the optimal control procedure adopted in our scheme. Our goal is to design the shape of the time-dependent optical tweezer fields that control the atom trajectories. We consider a system consisting of two atoms constrained to the plane $x-y$. The motion of the atoms is described by the classical Hamiltonian equations
	\begin{align}
		\dot{\mathbf{r}}^a =& \frac{\mathbf{p}^a}{m} \label{cl_Ham_sys_r}\\
		\dot{\mathbf{p}}^a =&	-\nabla_{\mathbf{r}^a} U + \mathbf{F}^{a} \quad a=1,2,\label{cl_Ham_sys_p}
	\end{align} 
	where $m$ is the particle mass, 
    $U$ is the potential generated by the optical tweezers, $	\mathbf{F}^{1}=-\mathbf{F}^{2}=- \langle	\Psi |\left(\nabla_{\mathbf{r}^1}\mathcal{H}_{dd}\right)\Psi\rangle $ is the dipole force, and $\mathcal{H}_{dd} (\mathbf{r}^1,\mathbf{r}^2)$ the dipole-dipole Hamiltonian. The two-atom wave function describes the internal states of the atoms and evolves according to the Schr\"odinger equation 
		\begin{align}
		i\dpp{\Psi}{t}= \mathcal{H}_{dd} (\mathbf{r}^1(t),\mathbf{r}^2(t))\Psi\;,
	\label{Sch}
	\end{align}
	where we have assumed the reduced Planck constant equal to one for simplicity. Typically, the dipole–dipole potential is the dominant interaction between two neutral Rydberg atoms, spatially separated. The dipole interaction of pairs of atoms in ultracold gases has been investigated theoretically and experimentally. By controlling the relative positions of two atoms in magneto-optical traps, it has been possible to measure both the intensity and the angular dependence of the dipole interaction in $^{87}$Rb atoms  \cite{Ravets_14,Ravet_15,Paris_18}. 
	The dipole-dipole Hamiltonian is given by $ \mathcal{H}_{dd} =   \frac{\mu}{|\mathbf{r}^2-\mathbf{r}^1|^3}\left(\mathbf{d}^1\cdot \mathbf{d}^2- 3 (\mathbf{d}^1\cdot \widehat{\mathbf{r}})(\mathbf{d}^2\cdot \widehat{\mathbf{r}}) \right)$, where $\mathbf{d}^a$ denotes the dipole momentum of the $a-$th atom, $\mu$ the interaction strength, and $\widehat{\mathbf{r}} = \frac{\mathbf{r}^2-\mathbf{r}^1}{|\mathbf{r}^2-\mathbf{r}^1|}$ the normalized relative position of the atoms \cite{Paris_18}. 
	We denote the internal state of the two-atom wave function by $|1,2\rangle=|j^1,m^2 \rangle\otimes |j^2,m^2\rangle$ where $|j^a,m^a\rangle$ specifies the orbital and spin quantum numbers of the $a-$th atom. 
    The matrix elements can be written as $
	\langle 1,2	| \mathcal{H}_{dd} |1',2'	\rangle =  \frac{\widetilde{\mu}_{1,1'}\widetilde{\mu}_{2,2'}}{|\mathbf{r}^2-\mathbf{r}^1|^3} \Big\langle 1,2	\Big| \mathcal{D} \Big|1',2'	\Big\rangle
	$, where $ \widetilde{\mu}_{1,1'}$, $\widetilde{\mu}_{2,2'}$ are denoted by radial factors. We have defined the operator
	\begin{align*}
		\mathcal{D} =&   (1-3\cos^2(\zq))d_{00}- \left(1-	\frac{3}{2}\sin^2(\zq) \right)\left(d_{+-}+d_{-+} \right) \\
		&-
		\frac{3}{2}\sin^2(\zq) \left( d_{++}+d_{--}  \right)-
		\frac{3}{\sqrt{2}}\sin(\zq)\cos(\zq)  \left(d_{-0}-d_{+0}  + d_{0-} -d_{0+}   \right)\;,
	\end{align*}
	and $d_{ij}\doteq \left(d_i \otimes d_j\right)$ with $i,j=0,\pm$. Here, the angle $\zq$ denotes the angle formed by the relative position of the atoms $\widehat{\mathbf{r}}$ with the quantization axis. 
		\begin{table}[!b]
		\begin{center}
			\begin{tabular}{ l  l} 
				Two-atom states\\[4pt]
					\hline 
				$| dd \rangle\doteq  
				\left| d,m=3/2;  d,m=3/2 \right\rangle $ &\quad  $| pf_1 \rangle\doteq   \left| p,m=1/2;  f,m=5/2 \right\rangle  $ \\[6pt]
				$| pf_2 \rangle\doteq  \left| p,m=1/2;  f,m=3/2 \right\rangle  $ &\quad $| pf_3 \rangle\doteq \left| p,m=1/2;  f,m=1/2 \right\rangle  $  \\ 
			\end{tabular}
			\caption{Two-particle atom states considered in our simulations. The $d$, $p$ and $f$ orbitals are $d=59D_{3/2}$, $p=61 P_{1/2}$, and $f=57F_{5/2}$.}\label{tab_orb}
		\end{center}
	\end{table}
	The matrix elements of the single-particle dipole operators are defined in terms of the Wigner 3-j symbols
	$
	\langle j, m| d_q |j', m'	\rangle = (-1)^{j'-1+m} \left(\begin{array}{ccc}
		j' & 1 & j\\
		m' & q & -m\\
	\end{array}\right)
	$
	with $q=0,\pm1$. In our simulations, we consider the four-dimensional Hilbert space spanned by the atom states $\{|dd\rangle,|pf_1\rangle,|pf_2\rangle,|pf_3\rangle \}$, for which the dipole interaction has been extensively investigated. The details of the atomic states are given in Tab. \ref{tab_orb}. 
%
	We assume the following simplification. We neglect the dependence of the radial coupling coefficient on the orbitals so that the dipole interaction strength reduces to a single parameter $ \widetilde{\mu}_{1,1'}\widetilde{\mu}_{2,2'}/h\doteq C_3/h=2.39$ GHz $\mu$m$^3$, where $h$ is the Planck constant \cite{Ravets_14,Ravets_15}. 
We pass to consider the effective field associated with the moving optical traps. The atoms are manipulated by two optical tweezers. One of the two is mobile and steers one atom along a closed path. In Fig. \ref{fig_res_cont_01}, we indicate by $A$ the initial position of the traveling atom. Using the second tweezer as a static trap, the second atom is held in a fixed position (point $B$). The optical fields of the mobile tweezer is modeled by a shifted Gaussian potential \cite{Endres_16}
	\begin{align}
	U_C(x,y;u_x,u_y) =U_D \, e^{- \frac{(x-u_x(t))^2 +(y-u_y(t))^2 }{\zs^2} } \; ,\label{tweez_pot}
	\end{align} 
where $(x,y)$ denote the spatial coordinates. The tweezer beam size $\zs$ and the potential width $U_D$ are considered constant. The tweezer field may be modified by varying the central position of the trapping potential $(u_x,u_y)$. This furnishes two control parameters that will be engineered by our optimal control procedure. The second static tweezer has a similar shape $U_T(x,y) =U_S \, e^{- \frac{(x-x_B)^2 +(y-y_B)^2 }{\zs^2} }$, where $(x_B,y_B)$ denote the coordinates of $B$.
 
Our goal is to design the shape of time-dependent control parameters performing the following tasks: $i)$ steer one atom along a closed trajectory while the second atom is held at rest; $ii)$ ensure that at the end of the procedure the wave function returns to the original state except for a phase factor; $iii)$ induce non-trivial geometric phase, with rejection of dynamic phase;  $iv)$ find initial and final two-particle states that can be expressed as a product of two single-atom states; $v)$ minimize the amount of energy required to control the system. All the previous requirements are modeled by a sum of suitable goal and cost functionals, whose minimum characterizes the optimal control of the system. 
In order to implement the $i)-iv)$ conditions,  we define the following goal functional 
	\begin{align*}
		\mathcal{G}=& 
		 \frac{1}{2}\sum_{a=1}^2\underbrace{\left(\chi_{\mathbf{r}} |\mathbf{r}^a(T)-\mathbf{r}^a(0)|^2 +\chi_{\mathbf{p}} |\mathbf{p}^a(T)|^2 \right)}_{i)}\\
		&-\underbrace{\chi_\Psi \left| \langle \Psi (0)| \Psi(T)\rangle \right|^2}_{ii)} 	+\underbrace{ \chi_{dy} \left| e^{\frac{i}{2}\zg_d }-1\right|^2}_{iii)} -\underbrace{ F(\Psi(T) )}_{iv)}\;.
	\end{align*} 
	The goal functional includes the set of non-negative  widths $(\chi_{\mathbf{r}},\chi_{\mathbf{p}},\chi_\Psi,\chi_{dy})$. They may be adjusted in order to fix the relative importance of the associated terms. The higher the width, the more the optimal solution will try to minimize the value of the corresponding term (see the discussion of similar cases in \cite{Morandi_24_OC_Wigner,Morandi_25_arxive}). As an example, the minimum of the first term of $\mathcal{G}$ is associated with optimal trajectories ending in proximity to the initial positions of the atoms $\mathbf{r}^a(0)$ and such that the atom velocity at the final time $T$ is as small as possible. The other terms have a similar interpretation.  
	The cost functional is given by
	\begin{align}
        \mathcal{K}=\frac{1}{2}\int_0^{T}\biggl[\zn_x\left|\dpt{u_x}{t}\right|^2+\zn_y\left|\dpt{u_y}{t}\right|^2 \biggr]\dif t \; ,
	\end{align}
	where $\nu_x,\nu_y >0$. The cost functional penalizes highly oscillating controls and ensures the stability of the minimization algorithm. The mathematical formulation of our optimal control problem consists in find 
	\begin{align}
	&\min_{u_x,u_y} \left\{\mathcal{G}+ \mathcal{K}\right\}    \textrm{, s. t. Eqs. } \eqref{cl_Ham_sys_r},\eqref{cl_Ham_sys_p},\eqref{Sch}\textrm{ hold true} \; .\label{opt_prob}
	\end{align}
	The optimal control of the two-atom system has been formulated as a constrained minimization problem. The Lagrangian formalism provides flexible tools for the mathematical implementation of optimal control protocols. Similarly to the standard Lagrangian multiplier technique, constrained problems are solved by defining additional unknowns associated to the constraints, in such a way that the original problem can be reformulated as an unconstrained minimization problem with respect to the total set of unknowns. Such additional unknowns, which constitute a generalization of the Lagrangian multiplier functions, are denoted by adjoint functions. We indicate by ${\mathbf{r}}^{ha},{\mathbf{p}}^{ha},\Phi$, the adjoint variables associated with the set of physical unknowns ${\mathbf{r}}^{a},{\mathbf{p}}^{a},\Psi$, respectively. 
	We define the following Lagrangian functional
	\begin{align*}
	\mathcal{L}	=& \mathcal{L}^{p-s}	+ \mathcal{L}^{\Psi}  + 	\mathcal{G}+ \mathcal{K}\;,
	\end{align*}
	where
	\begin{align*}
	\mathcal{L}^{p-s}	=& \sum_{a=1,2}\int_0^T \Big[\left( 	\dot{\mathbf{r}}^a - \frac{\mathbf{p}^a}{m} \right) \cdot \mathbf{p}^{ah} +  	\left(\dot{\mathbf{p}}^a +\nabla_{\mathbf{r}^a} U(\mathbf{u}) - \mathbf{F}^{ a}(\mathbf{r}^a,\Psi)\right) \cdot \mathbf{r}^{ah}  \Big] \dif t\\
	\mathcal{L}^{\Psi}	=& \int_0^T  2 \textrm{Re} \left[	 i \left(\Phi^{\dag} \dpp{\Psi}{t} \right) -\Phi^{\dag} \mathcal{H}_{dd}(\mathbf{r}^a) \Psi \right] \dif t\;.
	\end{align*}
	The Lagrangians $\mathcal{L}^{p-s} $ and $\mathcal{L}^{\Psi}$ correspond to the equations governing the evolution of the position and of the internal state of two atoms, respectively. The stationary points of the Lagrangian functional associated with the variations of  physical $({\mathbf{r}}^{a},{\mathbf{p}}^{a},\Psi)$ and  adjoint $({\mathbf{r}}^{ha},{\mathbf{p}}^{ha},\Phi)$ fields, constitute the solutions of the optimal control problem. For clarity,  we have indicated explicitly the dependence of the Lagrangian functionals $\mathcal{L}^{p-s} $ and $\mathcal{L}^{\Psi}$, with respect to the unknown fields. 
	The variation of the Lagrangian with respect to the adjoint phase-space variables provides the Hamiltonian Eqs. \eqref{cl_Ham_sys_r}-\eqref{cl_Ham_sys_p}. The Schr\"odinger Eq. \eqref{Sch} for the wave function $\Psi$ is recovered by imposing that $\mathcal{L}$ is stationary with respect to $\Psi$. The stationarity conditions $\delta_{\mathbf{r}^a} \mathcal{L}=0$, and $\delta_{\mathbf{p}^a} \mathcal{L}=0$, lead to the evolution equation for the adjoint trajectories 
	\begin{align*} 
		\dot{p}^{ah}_i =& \sum_{ j=x,y} \bigg(\frac{\partial^2 U}{\partial  r_{i}^a \partial r_{j}^{a }}- \sum_{a'=1,2 }\frac{\partial F_{j}^{a'd}}{\partial r_{i}^a } \bigg) r_{j}^{a'h} -2 \textrm{Re} \left(  \Phi^{\dag} \dpp{\mathcal{H}_{dd}}{r_i^a} \Psi \right)  \quad i=x,y\; ; a=1,2. 
		\\
		\dot{\mathbf{r}}^{ah} =&- \frac{\mathbf{p}^{ah}}{m} \;.
	\end{align*} 
	The evolution equations for the adjoint phase-space variables $(\mathbf{r}^{ah},\mathbf{p}^{ah})$ are completed by prescribing the coordinates at the final time $\mathbf{p}^{ah}(T)= \chi_{\mathbf{r}} (\mathbf{r}^{a}(0)-\mathbf{r}^{a}(T))$, and $\mathbf{r}^{ah}(T)= \chi_{\mathbf{p}} \mathbf{p}^{a}(T)$.
	The variation of $\mathcal{L}$ with respect to the field $\Psi^\dag$ provides the non homogeneous Schr\"odinger equation for the field $\Phi$
	\begin{align*} 
		i  \dpp{\Phi}{t} =& \mathcal{H}_{dd}
		\Phi-   \mathbf{p}^{ah}	\cdot \nabla_{\mathbf{r}^a}\mathcal{H}_{dd} \Psi  +\chi_{dy} \sin\left(\frac{\zg_g}{2}\right)	  \mathcal{H}_{dd}\Psi \;.
	\end{align*} 
	The Cauchy problem is completed by the final values conditions \begin{align*}
		\Phi(T)=&    i\chi_\Psi     \left( \Psi^\dag (0) \Psi(T) \right)\Psi(0)    +\frac{1}{2}\left(i\left.\dpp{F}{\textrm{Re}\Psi }\right|_{t=T}- \left.\dpp{F}{\textrm{Im}\Psi }\right|_{t=T} \right)\;.
	\end{align*}
	Finally, the control parameters are obtained by the stationarity of $\mathcal{L}$ with respect to $u_i$  
    \begin{align*}
	\nu_i \dst{u_i}{t} =\sum_{a=1,2;j=x,y} \frac{\partial^2 U}{\partial  r_j^a \partial u_i} p_j^{ah} &\quad  i=x,y\;.
	\end{align*}
The details concerning the derivation of the adjoint equations are given in Appendix \ref{App_der_adj_eq}. Further details can be found also in \cite{Morandi_24_OC_Wigner,Morandi_25_arxive,Morandi_24_OC_Bell}. For clarity, in Tab. \ref{tab_symbols} we list the symbols associated to the physical and adjoint fields and the parameters of the optimal control problem. 
\begin{table}[!ht]
	\begin{center}
		\begin{tabular}{ l  l} 
			Physical state \\
			\hline
			$(\mathbf{r}^{a},\mathbf{p}^{a})$, $a=1,2$ & Phase-space coordinates  \\
			$\Psi\in \mathbb{C}^4$ & Two-atom wave function  \\[4pt]			 
			Adjoint fields  \\
			\hline
			$(\mathbf{r}^{ah},\mathbf{p}^{ah})$, $a=1,2$ & Adjoint phase-space coordinates  \\
			$\Phi\in \mathbb{C}^4$ & Adjoint wave function  \\[4pt]			   
			Control fields  \\
			\hline
			$u_x,u_y$ & Center of the mobile optical tweezer  \\[4pt]
			Widths & Associated goal   \\
			\hline
			$\chi_{\mathbf{r}}$, $\chi_{\mathbf{p}}$, $\chi_\Psi$ & Ensure periodic motion: $\mathbf{r}^a(T)=\mathbf{r}^a(0)$, $\mathbf{p}^a(T)=0$, $\Psi(T)=\Psi(0)e^{i\za}$ \\
			 $\chi_{dy} $ &Minimize dynamical phase\\
			$\nu_x$, $\nu_y $& Minimize control cost        \\
		 \end{tabular}
		\caption{List of the relevant unknowns and parameters of the optimal control problem. }\label{tab_symbols}
	\end{center}
\end{table}
\section{Results}
	As indicated in the previous Section, our strategy consists in designing closed trajectories which are optimal in the sense of the criteria listed in Sec. \ref{sec_model}. 
	%
    In order to initialize the nonlinear minimization algorithm which provides the solution of the optimal control problem, it is necessary to find a starting approximation for the optimal trajectory. In this preliminary step, we require that the traveling atom follows a circle of radios $r$ and that the second is pinned in a fixed position located at the distance $d$ from the center of the circle. To obtain a first estimate of the optimal trajectories, we calculate the values of $r$ and $d$ that minimize the total cost functional $\mathcal{G}+\mathcal{K}$. After this initialization procedure is completed, we obtain the optimal trajectories of the tweezer and the atoms by solving the optimal control problem of Eq. \eqref{opt_prob}. 
    \begin{figure}[h!]
\includegraphics[width=0.95\textwidth]{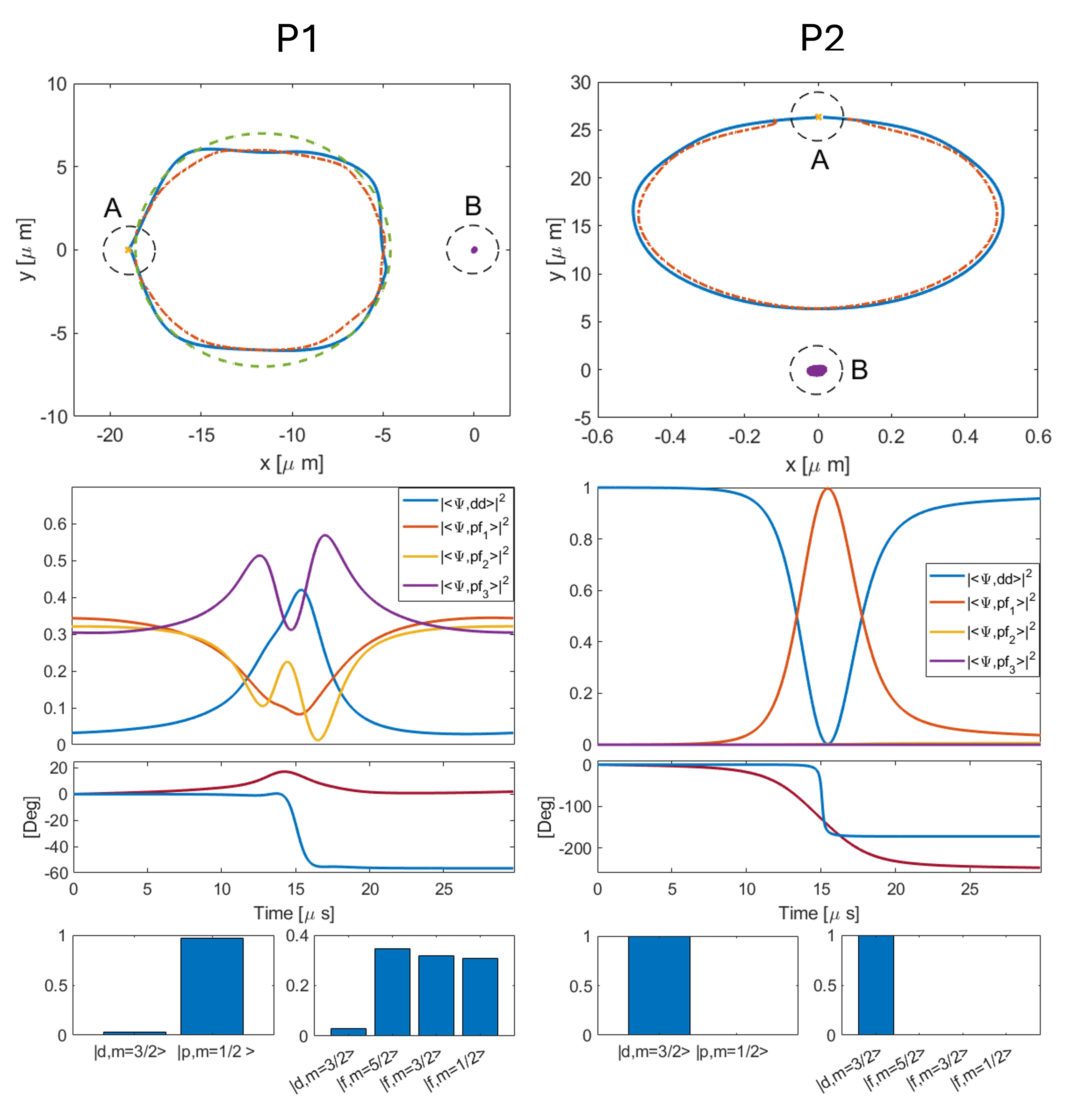}
	\caption{Control of the atom positions and AA phase. Top panel: trajectories of the atoms (blue and purple curves). The dashed red curves depict the trajectory of the center of the mobile tweezer. The points $A$ and $B$ denote the initial positions of the two atoms. Second row: evolution of the two-particle states occupation probabilities. Third row: time evolution of the geometrical (blue curve) and dynamical (red curve) phase. Button panels: single-particle occupation probabilities at the end of the process. The left subpanel refers to the traveling ant the right subpanel to the static atom. The left column of the figure refers to the P1 and the right column to the P2 case.  }\label{fig_res_cont_01}
\end{figure}

Due to the large number of degrees of freedom associated with the class of closed trajectories, several solutions of the optimal problem can be obtained. In particular, in the following we will discuss two solutions of the optimization problem which seem promising for experimental implementation. We will refer to such solutions as Protocol one (P1), and Protocol two (P2). In order to ease the comparison, we depict the two solutions in the same figure (Fig. \ref{fig_res_cont_01}), P1 in the left column and P2 in the right column. In our simulations, we use the following parameters: mobile tweezer depth $U_D=\qty{10}{\milli\kelvin}$, static tweezer depth $U_S=\qty{4}{\milli\kelvin}$, tweezers beam size $\zs=\qty{2}{\micro\meter}$,   simulation time interval final time $T=\qty{30}{\micro\second}$. 

We start by discussing in detail the solution P1. 
The panels in the first row depict the trajectories of the atoms in the plane $x-y$. The dashed circles around the points $A$ and $B$ are guides for the eyes and represent the initial and final positions of the two atoms. The distance between the initial point $A$ and the final point $B$ is fixed at $\qty{19}{\micro\meter}$. The trajectory of the moving particle is depicted by a continuous blue curve and, as required, starts and ends at the point $A$. The trajectory of the second particle is illustrated by a purple curve. We see that the second atom remains in the initial position $B$. The center of the tweezer that guides the traveling atom (curve $(u_x(t),u_y(t))$) is indicated by a red point-dashed curve.  We see that the atom is always located near the position of the minimum of the tweezer potential. The optimal trajectory deviates slightly from a circle. For comparison, we have depicted a circle (green dashed curve) of radius $\qty{7}{\micro\meter}$, whose center is at the distance of $d=\qty{11.6}{\micro\meter}$ from the point $B$. 
The occupation numbers of the two-particle wave function in the basis $\{| dd \rangle  ,| pf_1 \rangle  ,| pf_2 \rangle ,| pf_3 \rangle  \}$ are depicted in the second row of the left panels of Fig. \ref{fig_res_cont_01}. The plot shows that the occupation numbers are periodic in the evolution time interval $[0,T]$, with $T=\qty{30}{\micro\second}$, as expected. As indicated in Sec. \ref{Sec_mod_intro}, it is important to monitor the separability of the two-particle state, which is measured by the functional $F(\Psi(T))$. For P1 the separability is $F(\Psi(T))=99.2\%$ indicating that at the end of the loop the two-particle entanglement can be safely neglected. 
The time evolution of the dynamical (red curve) and geometrical (blue curve) phase are illustrated in the panels in the third row of Fig. \ref{fig_res_cont_01}. At the end of the path, the dynamical phase results to be small (around $\gamma^d\simeq 2^o$), and the geometric phase is $\gamma^g\simeq-56.7^o$. We note that the geometric phase varies significantly only in a small time interval centered at the middle of the path, where the atom distance reaches its minimum.
Finally, in the last row, we show the occupation numbers of the tow atoms in the single-particle basis at the end of the steering protocol. We see that the traveling atom is essentially in the $|p,m=1/2 \rangle $ state, while the second atom is in a linear superposition of the internal states $d$ and $f$. In order to observe the geometric phase associated to P1, the two atoms should be initially prepared according to the single-particle occupation probabilities illustrated in Figure. 	
By engineering optical pulses of lasers, it is nowadays possible to select internal states of Rydberg atoms with high fidelity. This finds application in quantum tomography, in the preparation of atomic states for quantum gates, and in the detection of quantum holonomy
\cite{Leseleuc_17,Sekiguchi_17,Zhou_17,Nagata_18,Saffman_20,Liu_20,Ai_20,Liang_22,Song_24}. As an example, in \cite{Leroux_18}, \isotope[87]{Sr} atoms are prepared in linear superposition of multiplet states by using a tripod scheme.
However, despite the fact that the optimization procedure has been successful in finding a path satisfying all the goals (closed loop, non-trivial geometric phase, vanishing dynamical phase) up to a satisfying degree, the fact that the second atom should be prepared as a linear superposition of single-particle states may lead to experimental challenges.

For this reason, we have developed a second protocol, denoted P2, where both atoms are prepared in a single internal state. P2 is illustrated in the right column of Fig. \ref{fig_res_cont_01}. As initial configuration, we have assumed that the atoms are aligned along the quantization direction (axes $y$) at a distance of $\qty{26.3}{\micro\meter}$. The trajectory of the traveling atom has the shape of an elongated ellipse. Similarly to the previous case, the particle trajectory (blue continuous curve) follows the minima of the tweezer field (red point-dashed curve).  The occupation numbers of the wave function $|\Psi\rangle $ are depicted in the second row. Compared to the previous case, we see that the evolution simplifies, being restricted to a single transition between the initial state $|dd\rangle $ and the auxiliary state $|pf_1\rangle$. The geometrical and dynamical phases are depicted in the panel in the third row and the final values are estimated as $\gamma_g\simeq= -172.5^o$ and $\gamma_{d}\simeq -247.8^o$, respectively. We see that the growth of the geometric phase is sharp and, as before, localizes halfway. In this case, it was not possible to reject efficiently the dynamical phase. Finally, the protocol indicates that both atoms should be prepared in the single state $|d,m=3/2 \rangle $ without significant superposition with different internal states.
  
\begin{figure}[h!]	
	\begin{center}		
		\includegraphics[width=0.49\textwidth]{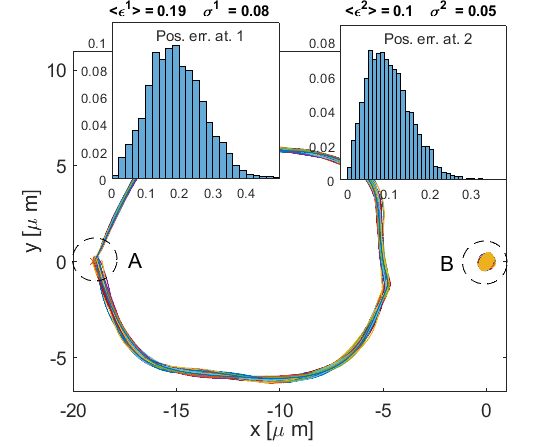}
		\includegraphics[width=0.5\textwidth]{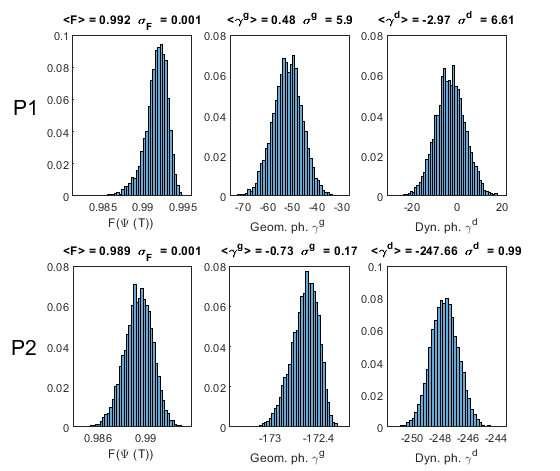}
		\caption{Left panel: atom trajectories perturbed by noise. We depict 100 realizations of Langevin-Wiener process for the case of P1. Left (right) inset depicts the statistics of the final position of the traveling (static) atom for $10^4$ realizations. $\langle\ze^i\rangle$  and $\zs^i$  with $i=1,2$ indicate the  mean and the standard deviation of the final position, respectively. Both quantities are expressed in $\qty{}{\micro\meter}$. Right panel: statistics associated to $10^4$ realizations. Upper panels refer to P1 and lower panels to P2 case. We depict, from the left to the right: purity functional $F(\Psi(T))$, geometric phase $\zg^g$, dynamic phase $\zg^d$. Mean and standard deviation of the distribution are indicated in the top of the panels.}\label{fig_res_noise_01}
	\end{center}
\end{figure}

Protocols concerning manipulations of ultracold atoms should consider the effects of experimental imperfections, the presence of thermal baths or sources of noise. Nowadays, trapped Rydberg atoms may be maintained to very low temperatures, typically below $\qty{}{\milli\kelvin}$, until few $\qty{}{\nano\kelvin}$. The main source of noise affecting optically trapped atoms arises from the statistical fluctuations of the laser fields. The laser fluctuations can be modeled by an effective stochastic force field. Similarly to collisions in Brownian motion, the stochastic noise interferes with the coherent transport of the atom driven by the optical tweezer and may degrade the precision of the protocol. We model the cumulative effects of the relevant sources of noise or thermal baths present in our system by a single Wiener-Langevin process. We modify the Hamilton equation for the momentum \eqref{cl_Ham_sys_p} by adding a Markovian stochastic force term with Gaussian probability distribution     \begin{align}
	\dot{\mathbf{p}}^a =&	-\nabla_{\mathbf{r}^a} U + \mathbf{F}^{a}-\zl \mathbf{p}^a+ {\bs \zh} \quad a=1,2,\label{cl_Ham_sys_p_noise}
\end{align} 
where the correlation of the white noise ${\bs \zh}$ has the Markovian form $\langle  \zh_i(t)\zh_j(t+\tau)\rangle= 2\zl k_B T \zd_{i,j}\zd(\tau)$, $T$ is an equivalent temperature modeling the combined effect of the external bath and the laser fluctuations, $\zl$ is the dumping coefficient, and $k_B$ the Boltzmann constant. In order to estimate the stability of our protocols, we run several realizations of the stochastic process and we recover the statistics of the physical quantities which are more relevant for the study of the geometric phase in atoms driven by tweezers. 
The results are depicted in Fig. \ref{fig_res_noise_01}. We compare the effect of noise on the two proposed implementations of the optimal control of AA phase, P1 and P2. We have run $N=10^4$ realizations of the Wiener stochastic process where we have fixed the bath temperature $T=\qty{0.1}{\milli\kelvin}$ and the momentum dumping coefficient $\zl=5\times 10^{-2}$ ms$^{-1}$ (corresponding to one ``collision'' every $\qty{20}{\milli\second}$). To illustrate the effect of the noise on the atom trajectories, in the upper panel, we have depicted the trajectories corresponding to 100 random realizations. In all the observed cases the tweezer field is able to steer correctly the first atom to the final position while the second atom undergoes to small fluctuations around the equilibrium position. In the insets, we depict the statistical distribution of the error $\ze^a = |\mathbf{r}^a(T)-\mathbf{r}^a(0)|$, where $a=1,2$, on the final position of the atoms  and we have indicated the relative standard deviation $\zs^a$. The left inset corresponds to the traveling atom and the right inset corresponds to the fixed atom. The plots illustrate the statistics of the complete set of $N=10^{4}$ realizations. We see that the protocol is stable and the final position error rarely exceed $\qty{0.4}{\nano\meter}$. In all the simulations, the algorithm had been successful in steering the atoms at the final position and no atom looses are observed. In the right panels, we compare the fluctuations of the separability and of the geometrical and dynamical phase of the two protocols. The first row corresponds to the protocol P1 and the second  row to the protocol P2. The simulations show that P2 proves to be clearly more robust than P1. For P1, the three observed variables (separability $F(\Psi(T))$,  geometric phase $\zg^g$, dynamic phase $\zg^d$) are practically unaffected by the noise, while for the protocol P1 we observe fluctuations in the range of around $\pm 7^o $ for both the geometric and the dynamic phase. The separability is stable and very high also for P1.     

\section{Conclusions}	

  In this work, a strategy to induce AA geometric phase in pairs of interacting atoms has been proposed. 
  The novelty of our proposal concerns the use of the mobile optical tweezers technology to induce two-particle correlation in a dynamical way, by modifying the relative position of the two atoms. The atom trajectories are shaped in such a way that the dipole-dipole interaction occurs in a controlled manner. 
  
  We have proposed two protocols with different characteristics. The first protocol has been successful in satisfying the goals that we have identified for the implementation of geometric phase in atom pairs: $i),ii)$ periodic evolution of the internal state apart from a global phase factor, $iii)$ rejection of the dynamical phase, $iv)$ separability of the two-atom  wave function at the beginning and at the end of the protocol. 
  Concerning the experimental implementation, the major limitations concern the preparation of the initial state and the sensitivity of the results to noise or experimental imperfections. In particular, the protocol requires the atoms to be prepared in a linear superposition of orbital states whose realization may be challenging. To avoid such difficulties, we have proposed a second protocol. In this case, both atoms are prepared in the same single orbital state. Moreover, up to temperature of $\qty{}{\milli\kelvin}$ the scheme is nearly insensitive to thermal noise. The disadvantage of this second scheme concerns the presence of a non-vanishing dynamic phase contribution to the total phase gained by the atom pairs.

\appendix 
\section*{Appendix}
\section*{Derivation of the adjoint equations}\label{App_der_adj_eq}
We consider
\begin{align*}
 \mathcal{L}^{\Psi}  +	\mathcal{G}^{\Psi}	=& \int_0^T \underbrace{2 \textrm{Re} \left[	 i \left(\Phi^{\dag} \dpp{\Psi}{t} \right) -\Phi^{\dag} \mathcal{H}_{dd} \Psi \right]}_{\mathcal{L}_{\Psi}^1}\dif t + \underbrace{\chi_\Psi \left| \langle \Psi (0)| \Psi(T)\rangle - e^{i\zg} \right|^2}_{\mathcal{L}_{\Psi}^2}  - F(\Psi)	\\
 &+ \chi_{dy} \left| e^{\frac{i}{2}\zg_g }-1\right|^2 ,
\end{align*}
where the dynamical phase is defined as $
\zg_d	= - \int_0^T  \left[	\Psi^{\dag} \mathcal{H}_{dd} \Psi \right]\dif t$. We have
	\begin{align*}
\delta_{\overline{\Psi}_{ij} }\mathcal{L}^1_{\Psi}	=& \delta_{\overline{\Psi}_{ij} }\int_0^T 2 \textrm{Re} \left[	- i \left( \dpp{\Phi^{\dag}}{t}\Psi \right) -\Phi^{\dag} \mathcal{H}_{dd} \Psi \right]_{0}^T
+ 2 \delta_{\overline{\Psi}_{ij} } \textrm{Re} \left[	 i \Phi^{\dag}\Psi  \right]_{0}^T\\
=& \delta_{\overline{\Psi}_{ij} }\int_0^T 2 \textrm{Re} \left[	 i \left( \Psi^{\dag} \dpp{\Phi}{t} \right) -\Psi^{\dag} \mathcal{H}_{dd}  \Phi\right]\dif t 
+ 2 \delta_{\overline{\Psi}_{ij} } \textrm{Re} \left[	- i \Psi^{\dag} \Phi \right]_{0}^T\\
=& \sum_{r,s}\int_0^T \left[	 i  \dpp{\Phi_{ij}}{t}  - [\mathcal{H}_{dd}]_{ij,rs}
  \Phi_{rs}\right]\delta {\overline{\Psi}_{ij} } \dif t 
	- i \Phi_{ij}(T)   \delta \overline{\Psi}_{ij}  (T)   +i  \Phi_{ij}(0)   \delta \overline{\Psi}_{ij}  (0) .
\end{align*}
Here, overline denotes conjugation. Furthermore, 
	\begin{align*}	
\delta_{\overline{\Psi}_{ij} }	\mathbf{F}^{a}= \sum_{r,s}\delta_{\overline{\Psi}_{ij} } \left[-\overline{\Psi}_{ij} 	[\nabla_{\mathbf{r}^a}\mathcal{H}_{dd}]_{ij,rs}\Psi_{rs} \right]=\sum_{r,s}
  \left[- 	[\nabla_{\mathbf{r}^a}\mathcal{H}_{dd}]_{ij,rs}\Psi_{rs} \right]\delta\overline{\Psi}_{ij} =
  - 	[\nabla_{\mathbf{r}^a}\mathcal{H}_{dd}\Psi]_{ij} \;\delta\overline{\Psi}_{ij} ,
\end{align*}
and
\begin{align*}
	\delta_{\overline{\Psi}_{ij} }\mathcal{L}^2_{\Psi}	=& \chi_\Psi \delta_{\overline{\Psi}_{ij} } \left( \Psi (0)^{\dag} \Psi(T) - e^{i\zg} \right)\left( \Psi(T)^{\dag} \Psi (0) - e^{-i\zg} \right)\\
	=& \chi_\Psi \left(   \Psi_{ij}(T)\delta \overline{\Psi_{ij}} (0)  -i e^{i\zg}  \delta_{\overline{\Psi}_{ij}} \zg \right)\left( \Psi(T)^{\dag} \Psi (0) - e^{-i\zg} \right)\\
	&+\chi_\Psi\left( \Psi (0)^{\dag} \Psi(T) - e^{i\zg} \right)\left(  \Psi_{ij} (0) \delta\overline{\Psi}(T)_{ij} +i  e^{-i\zg}\delta_{\overline{\Psi}_{ij}} \zg  \right),
\end{align*}
where $
	 \delta_{\overline{\Psi}_{ij}} \zg =-\int_{0}^T [\mathcal{H}_{dd}\Psi]_{ij} \delta{\overline{\Psi}_{ij}}\dif t
$. We have
	\begin{align*}
	\delta_{\overline{\Psi}_{ij} }	\tilde{\mathcal{L}}_{\Psi}^2 =&  -	\chi_\Psi	\delta_{\overline{\Psi}_{ij} }  \left|  \Psi^\dag (0) \Psi(T) \right|^2\\
	&=  -\chi_\Psi  \left( 	 \Psi_{ij}(T)  \left( \Psi^\dag (T) \Psi(0) \right) \delta{\overline{\Psi}_{ij} } (0) +  \left( \Psi^\dag (0) \Psi(T) \right)\Psi_{ij}(0)  \delta{\overline{\Psi}_{ij} } (T)  \right),
	\end{align*} 
and
	\begin{align*}	
	\delta_{\overline{\Psi}_{ij} }	F (\Psi(T))=	\dpp{F(\Psi(T))}{\overline{\Psi}_{ij}} \delta {\overline{\Psi}_{ij} (T)}=\frac{1}{2}\left(\left.\dpp{F}{\textrm{Re}\Psi_{ij} }\right|_{t=T}+i \left.\dpp{F}{\textrm{Im}\Psi_{ij} }\right|_{t=T} \right)\delta \overline{\Psi}_{ij} (T).
\end{align*}
Concerning the dynamical phase we have
\begin{align*}	
\delta_{\overline{\Psi}_{ij} }	\left| e^{\frac{i}{2}\zg_g }-1\right|^2 =\sin\left(\frac{\zg_g}{2}\right)\delta_{\overline{\Psi}_{ij} }\zg_g =-
\sin\left(\frac{\zg_g}{2}\right)	[\mathcal{H}_{dd}\Psi]_{ij} \;\delta\overline{\Psi}_{ij},
\end{align*}
and we obtain the following equation for the adjoint wave function
\begin{align*}
&  i  \dpp{\Phi}{t} = \mathcal{H}_{dd}
	\Phi-   \mathbf{p}^{ah}	\cdot \nabla_{\mathbf{r}^a}\mathcal{H}_{dd} \Psi  +\chi_{dy} \sin\left(\frac{\zg_g}{2}\right)	  \mathcal{H}_{dd}\Psi, 
\end{align*}
with final condition
\begin{align*}
\Phi_{ij}(T)=&    i\chi_\Psi     \left( \Psi^\dag (0) \Psi(T) \right)\Psi_{ij}(0)    +\frac{1}{2}\left(i\left.\dpp{F}{\textrm{Re}\Psi_{ij} }\right|_{t=T}- \left.\dpp{F}{\textrm{Im}\Psi_{ij} }\right|_{t=T} \right).
\end{align*}

\begin{acknowledgments}
The work has been developed under the auspices of GNFM (INdAM). \end{acknowledgments}

	



\end{document}